**Improvement of critical current density in the $Cu/MgB_2$ and $Ag/MgB_2$ superconducting wires using the fast formation method**

S. Soltanian, X.L. Wang*, J. Horvat, A. H. Li, H.K. Liu and S.X. Dou

Institute for Superconducting and Electronic Materials, University of Wollongong, Northfields Ave, NSW 2522 Australia

## Abstract

The powder in tube method has been used to fabricate Ag and Cu clad $MgB_2$ wires using an *in-situ* reaction method. The effects of short time sintering on the critical current densities of Ag and Cu clad $MgB_2$ wires were studied. All the samples were examined using XRD, SEM, and magnetization measurements. For Ag clad wire Jc is improved by more than two times after the short time sintering process. $J_c$ values of $1.2\times10^5$ A/cm$^2$ in zero field and above $10^4$ A/cm$^2$ in 2T at 20 K have been achieved for Ag clad $MgB_2$ wire which is only sintered for 6 minutes at 800$^o$C. However, a remarkable degree of reaction has been found between the superconducting cores and the sheath materials, leading to the formation of $Cu_2Mg$ and $Ag_3Mg$ for copper and silver clad wires, respectively. The results from $T_c$, $J_c$ and $H_{irr}$ convincingly show that the short sintering causes less reaction between the magnesium and the sheath materials and markedly improves the critical current density. Our result shows that Iron is still the best sheath material because of the lack of reaction between Fe and the superconducting $MgB_2$ material.

*Corresponding author. Tel.: +61-2-422-15766; Fax: +61-2-422-15731.
Email address: xiaolin@uow.edu.au (X. L. Wang)

## Introduction

The recent discovery of superconductivity in $MgB_2$ with a transition temperature of 39K by Akimitsu et al. [1] has led to great progress researches in applied superconductivity. Unlike the case of copper oxide superconductors it has been proved that Jc in $MgB_2$ is not limited by weak links.[2,3] To achieve high critical current density, different techniques have been developed [4-15]. Among these techniques, the powder-in-tube (PIT) method appears most promising and practical. Some metals and alloys have been found to be suitable for sheath materials in the PIT process. Iron and its alloys in particular have been found to be not only non-poisoning to $MgB_2$ [4,7-11] but also capable of providing magnetic screening to reduce the effect of external applied magnetic fields on the critical currents [9,11]. A high transport $J_c$ on the order of $10^4$-$10^5$ A/cm$^2$ at 20 K and 4.2 K has been reported for Cu/Fe/$MgB_2$ tapes where reacted $MgB_2$ powders were used as the core conductor and sintered at 900-1000$^o$C for 0.5 h out of a total heat treatment time of more than 3 h, including the initial heating [4]. By using unreacted Mg+2B powders and sintering at 800$^o$C for 1 h, Fe clad $MgB_2$ tapes with a high transport $J_c$ above $10^4$ A/cm$^2$ at 30 K and 1 Tesla and $I_c$ greater than 150 A have been successfully fabricated by our group [9]. Recently Fe and Ni clad wire have been fabricated with quite high Jc values of 2.3x$10^5$ A/cm$^2$ at 4.2K and 1.5T in a Ni/$MgB_2$ tape and $10^4$ A/cm$^2$ at 4.2K and 6.5T in a Fe/$MgB_2$ [12].

Although the high toughness materials have some benefits in achieving high-density samples, these materials are usually very hard to mechanically deform. They are also easily broken. So, nonmagnetic and easily deformable silver and copper can be the best alternative for sheath materials, especially for some applications such as superconducting magnets. Ag and Cu-clad $MgB_2$ tapes using in-situ and ex-situ reactions have been already prepared [14]. A magnetic Jc value of above $10^4$ A/cm$^2$ has been reported By Glowacki et al. at 5K and low magnetic field for Cu clad $MgB_2$ wire, which was sintered at 620°C for 48h [15]. More recently they have also reported relatively higher Jc values for Cu clad $MgB_2$ wire that was heat-treated at 700°C for 1h [16]. Very recently, Wang et al. found that the $MgB_2$ superconducting phase can be formed in a very short time. Just a few minutes of heat treatment is enough to achieve high quality Fe/$MgB_2$ wires with a high $J_c$ value of $4.5 \times 10^5$ A/cm$^2$ at 15K and 1T [8].

As it has been reported that Ag and Cu react with Mg [14,16], it is proposed that a shorter sintering time would decrease the reaction of Mg with Ag and Cu sheath materials and lead to some improvements in wire performance. In this paper we have investigated the effect of short time sintering on the Jc of Ag and Cu/$MgB_2$ wires and compared these results with the performance of Fe/$MgB_2$ wire that also prepared with the short time sintering.

## Experimental Details

Cu and Ag-clad $MgB_2$ wires were prepared using a standard powder-in-tube (PIT) method. Powders of magnesium (99%) and amorphous boron (99%) with the stoichiometry of $MgB_2$ were well mixed. The mixed powder was then loaded into Cu and Ag tubes. The pure Ag and Cu tubes had an outside diameter (OD) of 8 mm, a wall thickness of 1 mm, and were 10 cm long. One end of the tube was sealed. The tubes were filled in with mixed $MgB_2$ powders. The remaining ends were crimped mechanically. Each tube was drawn to a wire about 1.5 mm in diameter and 2 meters long. The Fe-clad $MgB_2$ wire preparation has been explained elsewhere [8,9].

Short length Ag and Cu-clad wire samples about 2 cm in length were sintered using the fast formation method [8]. One of the Ag-clad samples was also sintered by the normal longer sintering.

For fast sintering, Ag and Cu-clad wire samples are sealed in a small Fe tube and then directly heated at a preset temperature of 800$^{o}$C for 6 minutes in flowing high purity Ar. This is then followed by a quench in liquid nitrogen. In the normal sintering case, one Ag-clad $MgB_2$ sample was sealed in a small Fe tube and then sintered in a sealed tube furnace in flowing high purity Ar. In this case, the temperature was increased at a heating rate of 600$^{o}$C/h to 800$^{o}$C, then furnace cooled down to room temperature without any holding period at 800$^{o}$C.

Fig. 1 shows the real temperature of the normally sintered sample as a function of time. The inset shows the average temperature variation of the fast sintered samples with time, starting from when the samples were loaded into a hot tube furnace held at a constant temperature of 800$^{o}$C.

Fig. 1 reveals that for the normal sintering case, the sample experienced temperatures higher than 660$^{o}$C, which is the magnesium melting point, for about 77 minutes. So we called this sample as the long time-sintered (LS) sample. However, for the fast formation case, this period is only a few minutes (about 4.5 min.) and we call these samples short-time sintered (SS) samples. The surface of the Fe tubes used to seal the wires was slightly oxidized after sintering. However, the $MgB_2$ wire samples sealed inside the Fe tubes were as fresh as before sintering.

The phase and crystal structure of all the samples were obtained from X-ray diffraction patterns using a Philips (PW1730) with Cu$K\alpha$ radiation. By opening the wires and removing the superconductor cores mechanically, X-ray diffraction (XRD) of the internal surface of the Ag and Cu sheath materials can be performed.

Microstructure study and surface analysis were carried out using the scanning electron microscope (SEM). The dc field dependence of the magnetization was measured using the Physical Property Measurement System (PPMS), Quantum Design® between 5K and 35K at different dc fields up to 6T. For magnetic characterization of Fe clad wire, bare cores were used because of the strong shielding effect of the Fe sheath metal [10]. Cylindrical bars of $MgB_2$ core were obtained by mechanically removing the Fe sheath.

**Result and Discussion**

XRD diffraction patterns of $MgB_2$ core, separated from the Ag and Cu-clad wires are shown in Fig. 2. It can be seen that both revealed relatively single $MgB_2$ phase with a slight amount of MgO (<5%). In order to study the reaction between the core and the sheath material at the interface, XRD diffraction patterns of the internal surface of the Ag and Cu were done after the cores were removed mechanically. The XRD results are shown in Fig. 3. It shows that magnesium has reacted with the sheath materials and formed $Cu_2Mg$ and $Ag_3Mg$ phase on the internal sheath surface of the Cu and Ag clad wires, respectively. The unknown peaks that were not matched with PDF database lines are indicated by question marks. More work is in progress to identify these peaks. For Fe clad wire there was not any clear evidence for any reaction between Mg or B and Fe. These results are in agreement with recent reports [14,16].

Fig. 4 presents a magnified view of a transverse cross section for Cu/$MgB_2$ wire using SEM back scattered electron imaging. A well-defined reacted layer with about 40μm in thickness can be clearly seen. It is due to the reaction of magnesium and copper at high temperature in agreement with the XRD diffraction pattern (Fig. 3). The diffusion of magnesium into the copper sheath is clear in Fig. 5, which shows the EDS surface analysis result. The magnesium concentration in the central part of the superconductor core is higher than in the area close to the copper sheath. The reacted layer in the SS Ag clad wire is also about 25μm (Fig. 6). However as we can see in Fig. 7, the reacted layer in the LS Ag clad wire is much thicker, about 90μm. So increasing the sintering time increases the magnesium deficiency and consequently causes a deficiency $MgB_2$ phase in the superconducting core. Longer sintering times could thus lead to a lower $J_c$ in the wire.

Measurements of the M-H loops at different temperatures were carried out on the Cu, Fe and Ag-clad wires. A typical M-H loop of an Ag-clad SS wire sample is shown in Fig. 8. We can see that a typical flux-jumping pattern is present for temperatures below 15 K. This flux jumping has been also observed in $MgB_2$ bulk samples [17].

The critical current density was calculated from the M-H loops using the Bean critical model $J_c = 30\ \Delta M/d$, where $\Delta M$ is the height of the M-H loop, and d is the diameter of the cylinder of the bare core. The field dependence of $J_c$ up to 6 Tesla was measured for three samples of Cu and Ag-clad wires at 5 K, 20K and 30 K as are shown in Fig. 9. The field dependence of Fe clad wire at 20K and 30 K is also shown. As we can see, Fe-clad wire has the highest Jc and the best Jc-field dependence among all samples at 20K and 30K. It should be noted that a $J_c$ of $1.3\times10^5$ A/cm$^2$ at 20K and zero field has been achieved for the Ag-clad SS sample. It is not possible to exactly measure the $J_c$ at low fields at temperatures below 15K due to the flux jumping. Ag-clad LS wire has the lowest $J_c$ at low fields for all entire temperature range. Although the Cu-clad wire has a higher $J_c$ than the Ag-clad LS wire at low field, the $J_c$-field performance of this sample is not as good as for the Ag-clad LS sample. The Ag-clad LS sample has slightly higher $J_c$ than Cu-clad wire at high fields over the entire temperature range probably due to poor grain connectivity in the Cu-clad wire.

The field dependence of $J_c$ for the SS copper wire was compared with the recent results which were reported by Glowacki et al. [15,16], as is shown in Fig. 10. As we can see, the $J_c$ of wire that was sintered at 700$^o$C for 1h is about two times higher than for wire that was sintered at 620$^o$C for 48h. Our copper wire, which was sintered at 800$^o$C for 6 min, has better $J_c$-field performance than wire that was sintered at 620$^o$C for 48h. It even has better $J_c$-field performance at low field (less than 4Tesla) than wire that was sintered at 700$^o$C for 1h. The slightly inferior performance of our SS wire at high fields is probably because the longer sintering time in the 700$^o$C sample caused stronger grain connectivity. The $J_c$ field dependence for short time (SS) and long time sintered (LS) Ag wires at 10K are shown in the inset. As we can see the $J_c$ in the SS sample is more than two times higher than for the LS sample due to less reaction between the superconducting core and sheath material.

Fig. 11 shows the irreversibility fields ($H_{irr}$) versus temperature for all the samples. $H_{irr}$ was determined from $J_c$ – H curves using the criterion of 100 A/cm$^2$. We can see that the copper clad wire has the lowest $H_{irr}$ for the whole temperature range. The SS Ag-clad wire also has a higher $H_{irr}$ than the LS wire over the whole temperature range. Fe has the highest $H_{irr}$ among all the samples. As we can see, the differences between the $H_{irr}$ values are increased by decreasing the temperature.

## Conclusion

In this paper we have investigated the effects of sintering time and temperature on the critical current densities of Cu, Ag and Fe-clad $MgB_2$ wires. It was found that a short time heat treatment in the fabrication of Cu and Ag clad $MgB_2$ wires can markedly enhance the critical current density. A total sintering time of several minutes is enough to form nearly pure $MgB_2$ with high performance characteristics. The $T_c$, $J_c$ and $H_{irr}$ results show that the Cu and Ag clad $MgB_2$ wires samples which were sintered for 6 minutes are better than those sintered for longer times. $J_c$ of $1.2\times10^5$ A/cm$^2$ in zero field and above $10^4$ A/cm$^2$ in 2T at 20 K have been achieved for the Ag clad $MgB_2$ wire.

Acknowledgement: The authors thank Dr. T. Silver for her helpful comments and discussion. S. Soltanian thanks the Department of Physics, University of Kurdistan, Iran for providing financial support for his Ph.D study at the University of Wollongong. This work was supported by funding from the Australian Research Council.

**Figure Captions:**

Fig.1. Real temperature that sample has experienced as a function of time for normal sintered sample. The inset shows the time variation of the average real temperature of the short sintered sample, starting from when the wires were loaded into a hot tube furnace held at a constant temperature of $800^{o}$C.

Fig. 2. XRD patterns recorded from the superconducting core of one of the samples when the Ag and Cu sheath materials were mechanically removed.

Fig. 3. XRD patterns recorded from the internal surface of the sheath of Ag and Cu-clad MgB2 wire samples when the superconducting core was mechanically removed.

Fig. 4. Scanning Electron Microscope image for a transverse cross-section of Cu/MgB2 sample using back scattered electron imaging.

Fig. 5. Scanning Electron Microscope image and EDS surface analysis for a transverse cross-section of Cu/MgB2 sample.

Fig. 6. Scanning Electron Microscope image for the transverse cross-section of the long time sintered Ag/MgB2 sample using back scattered electron imaging.

Fig. 7. Scanning Electron Microscope image for a transverse cross-section of the short time sintered Ag/MgB2 sample using back scattered electron imaging.

Fig. 8. M-H loop of long time sintered Ag/$MgB_2$ sample at different temperatures.

Fig. 9. Field dependence of $J_c$ at 5K, 20K and 30K for Ag, Cu and Fe clad wires.

Fig. 10. The comparison between $J_c$ field dependence of our Cu clad wire and the Cu clad wires, which were reported by Glowacki et al. [16,17] at 5K. The $J_c$ field dependence for short and long sintered Ag wires at 10K is shown in the inset.

Fig. 11. Irreversibility lines for all the samples.

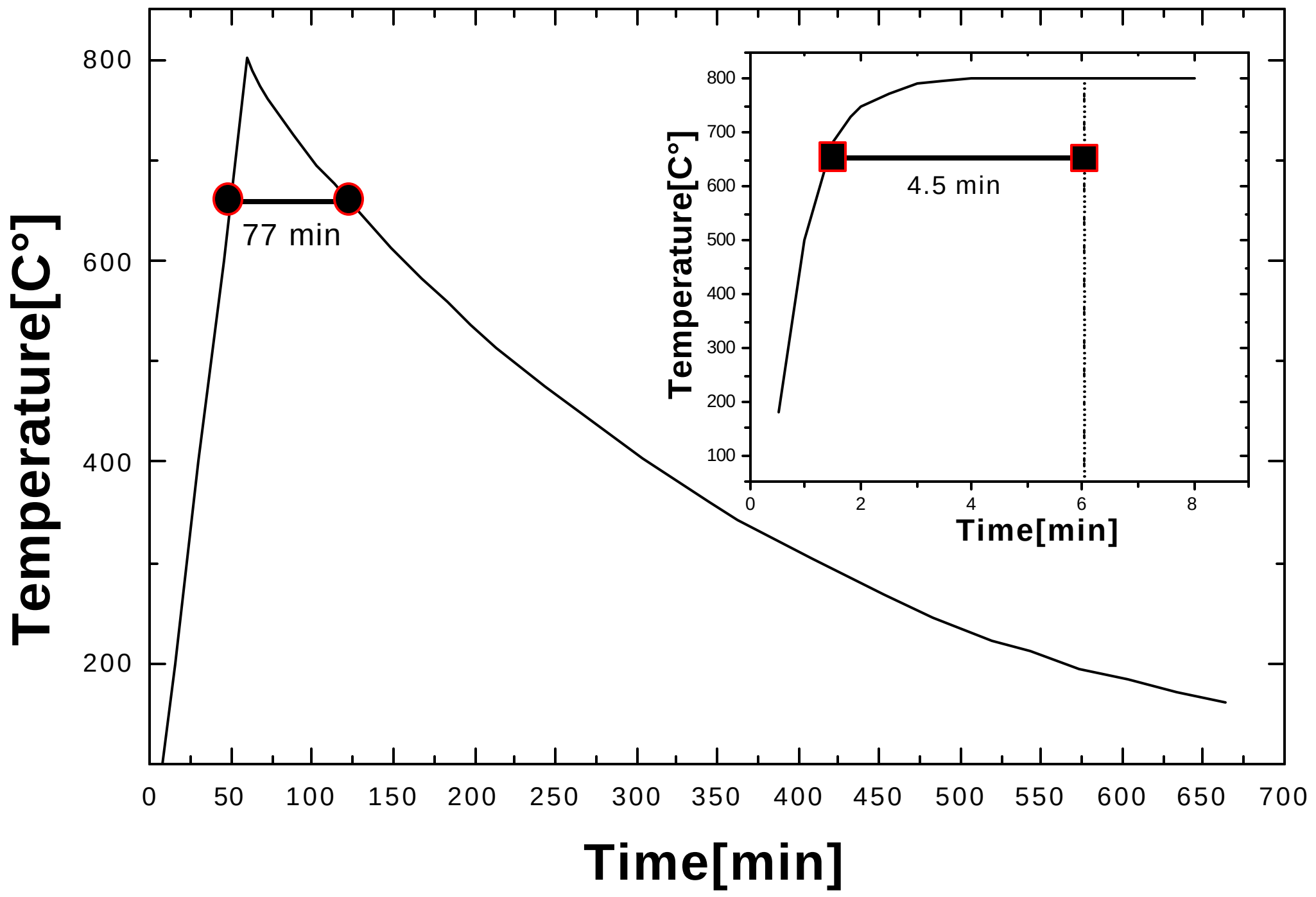

Fig. 1

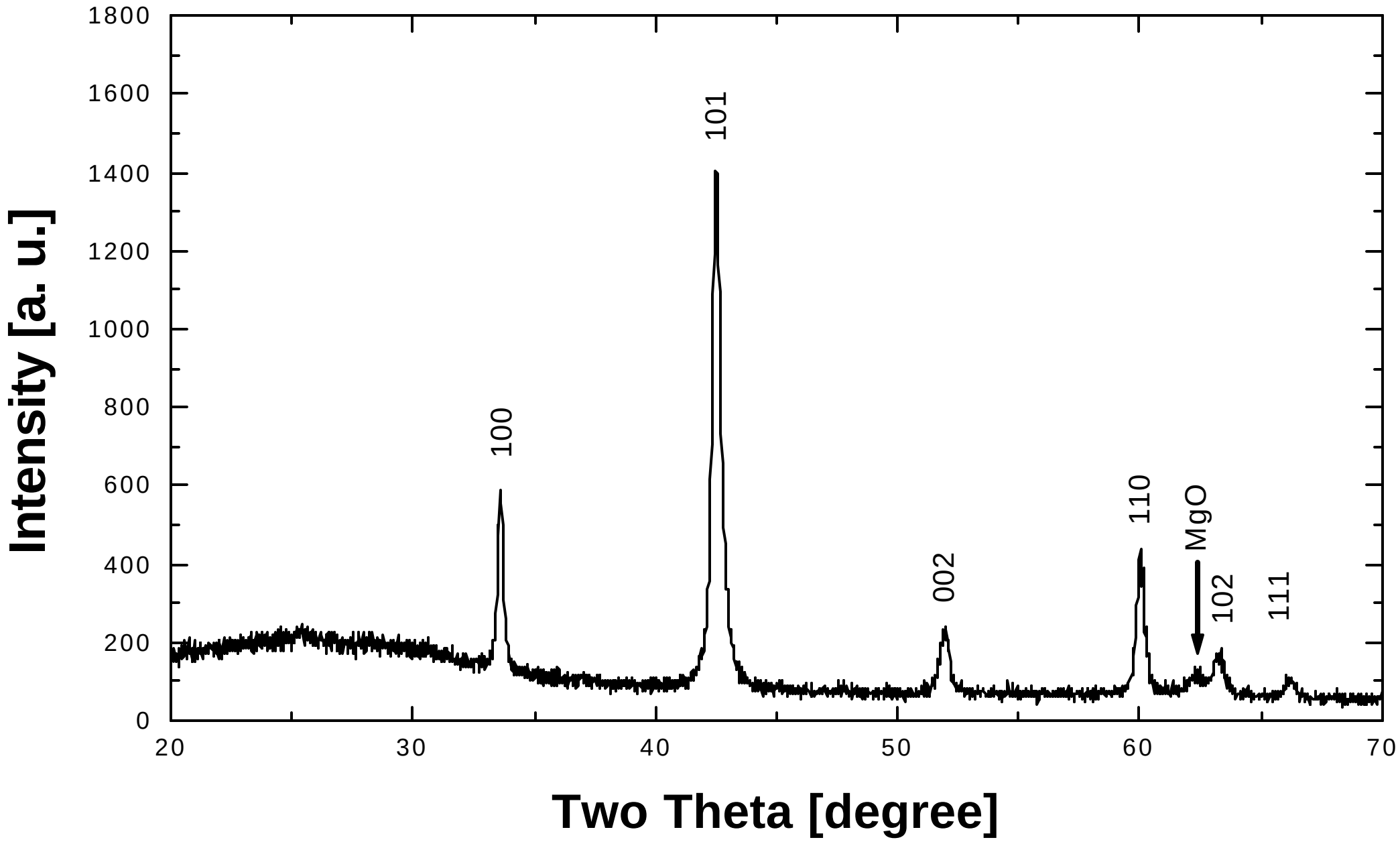

Fig. 2

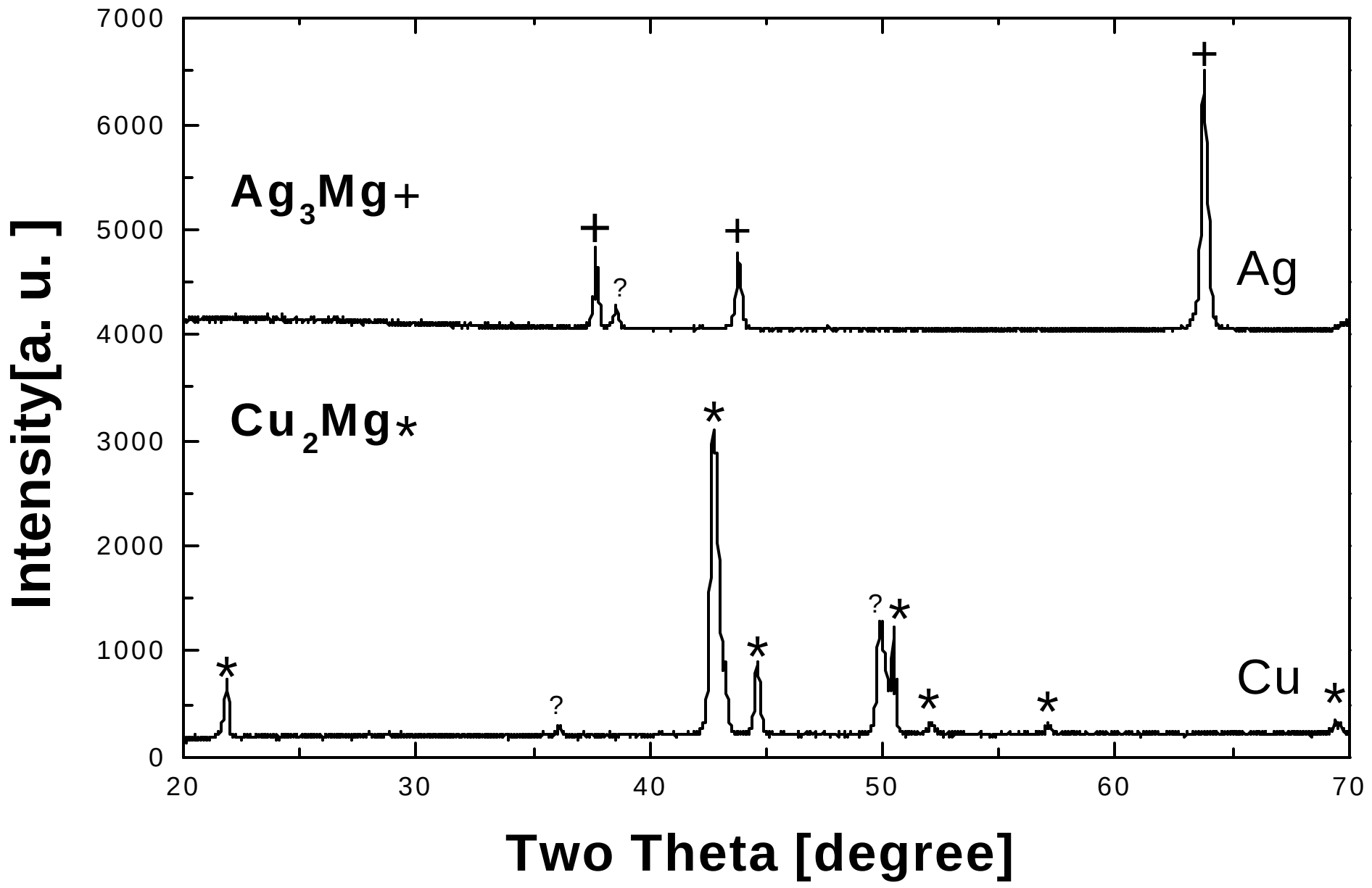

Fig. 3

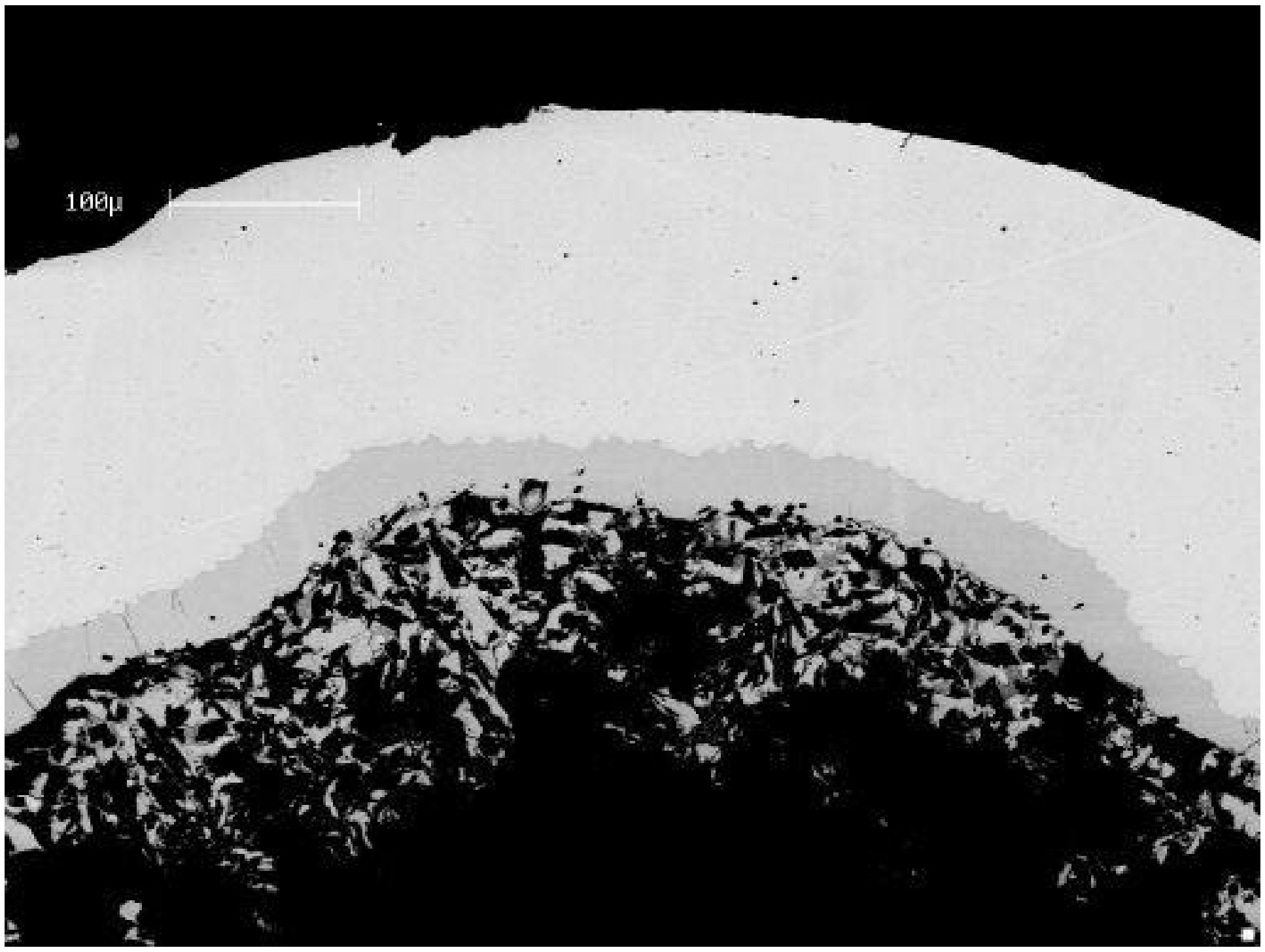
100µ

Fig. 4

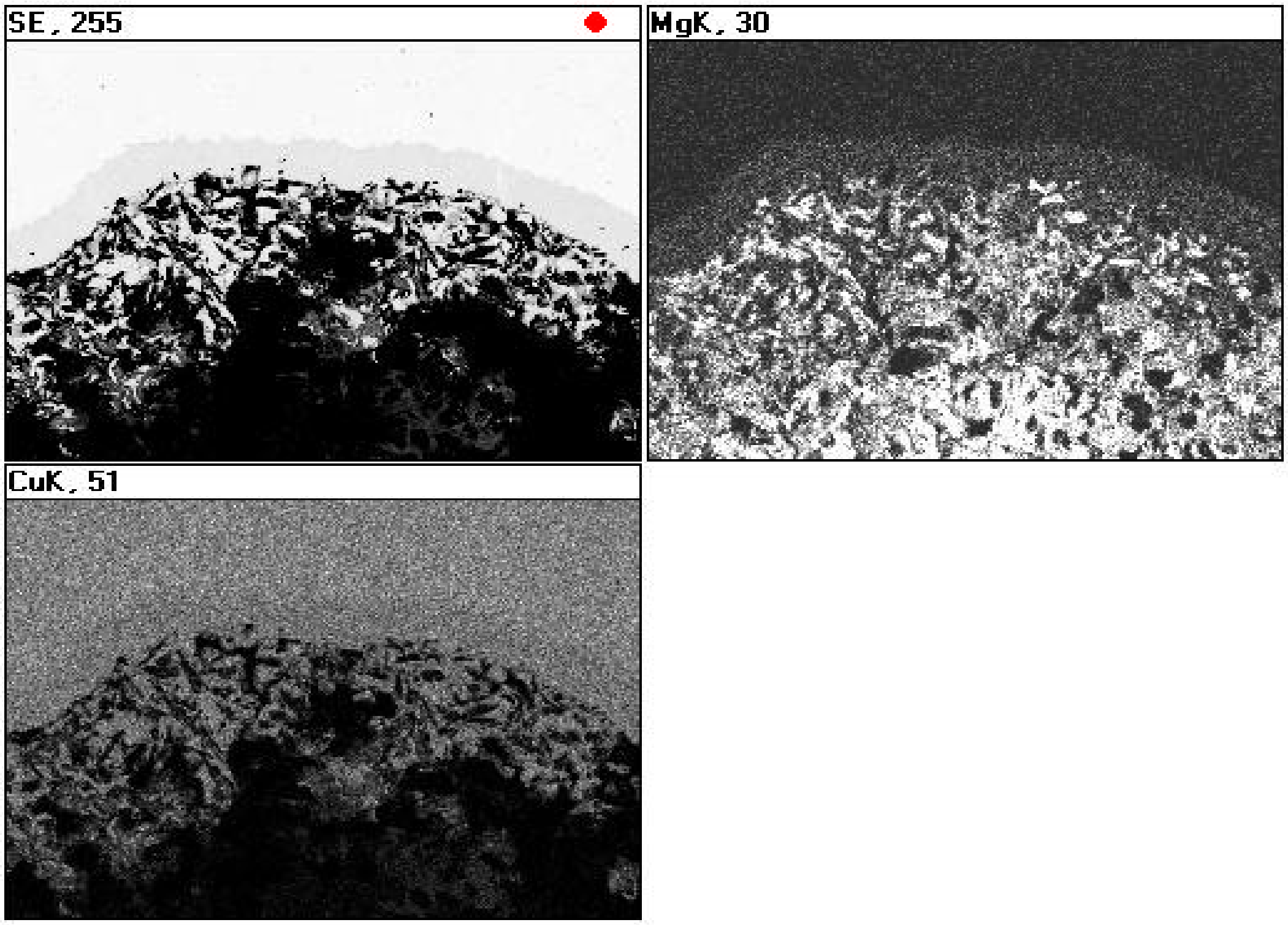

Fig. 5

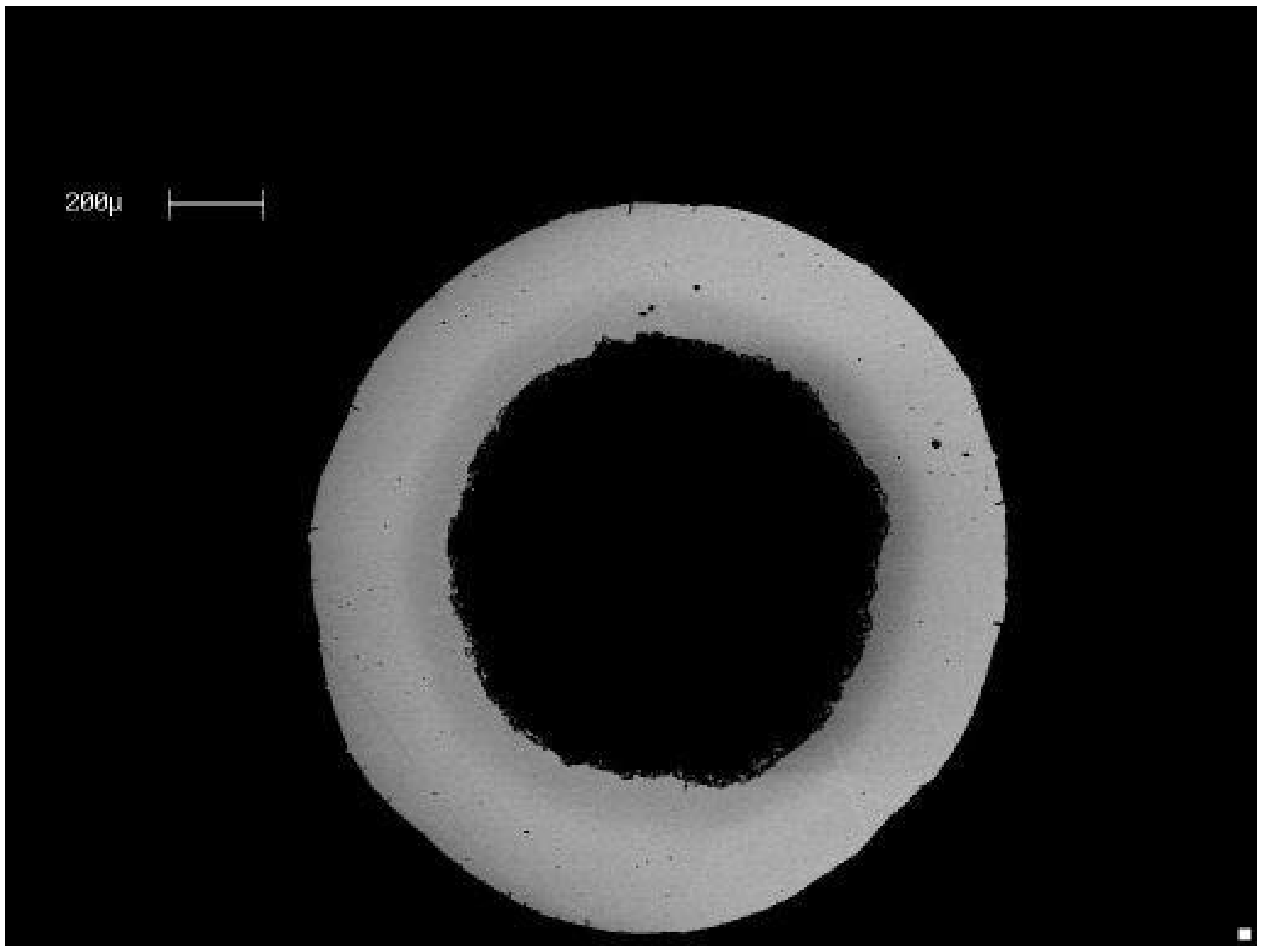

Fig. 6

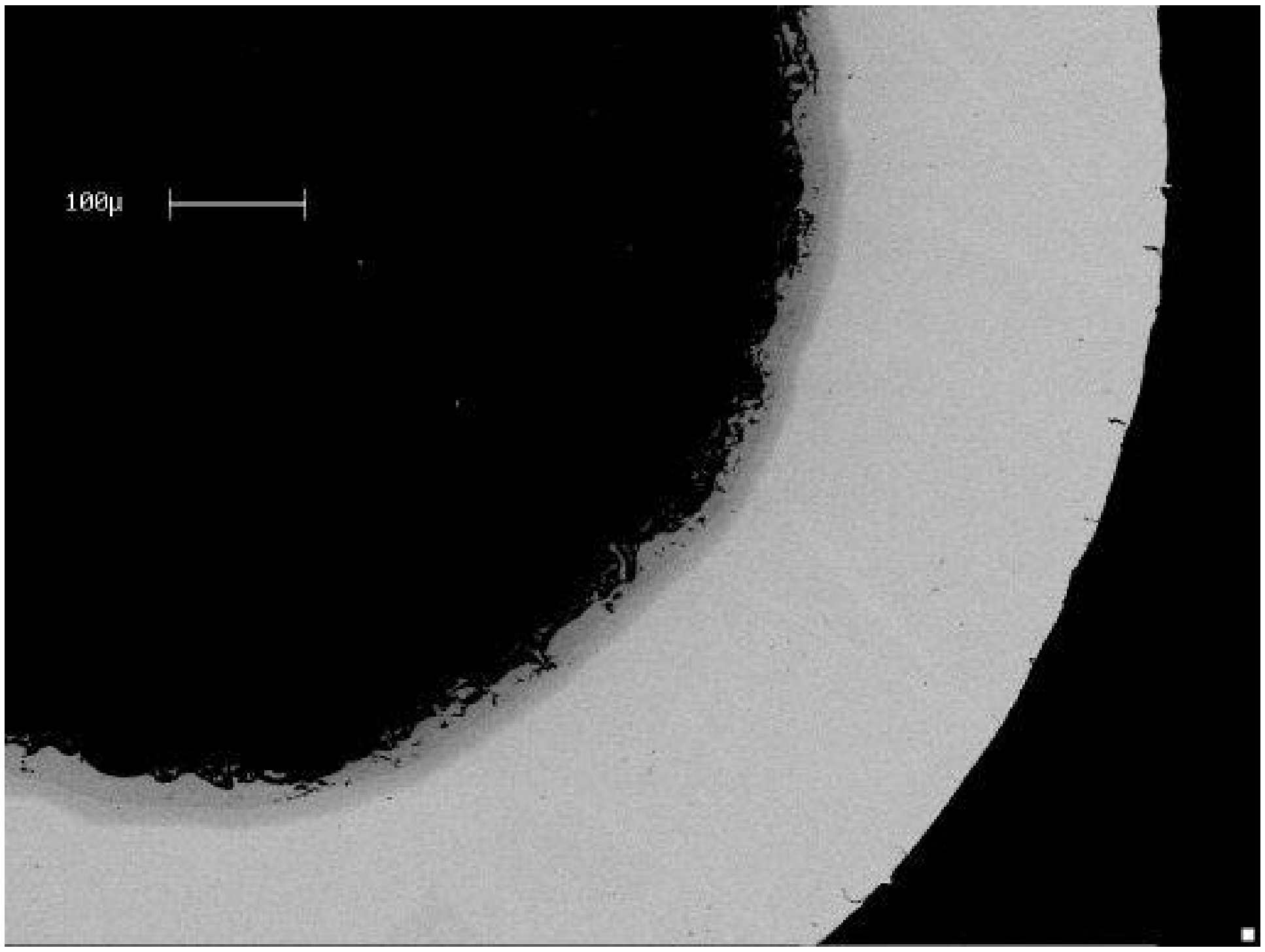

Fig. 7

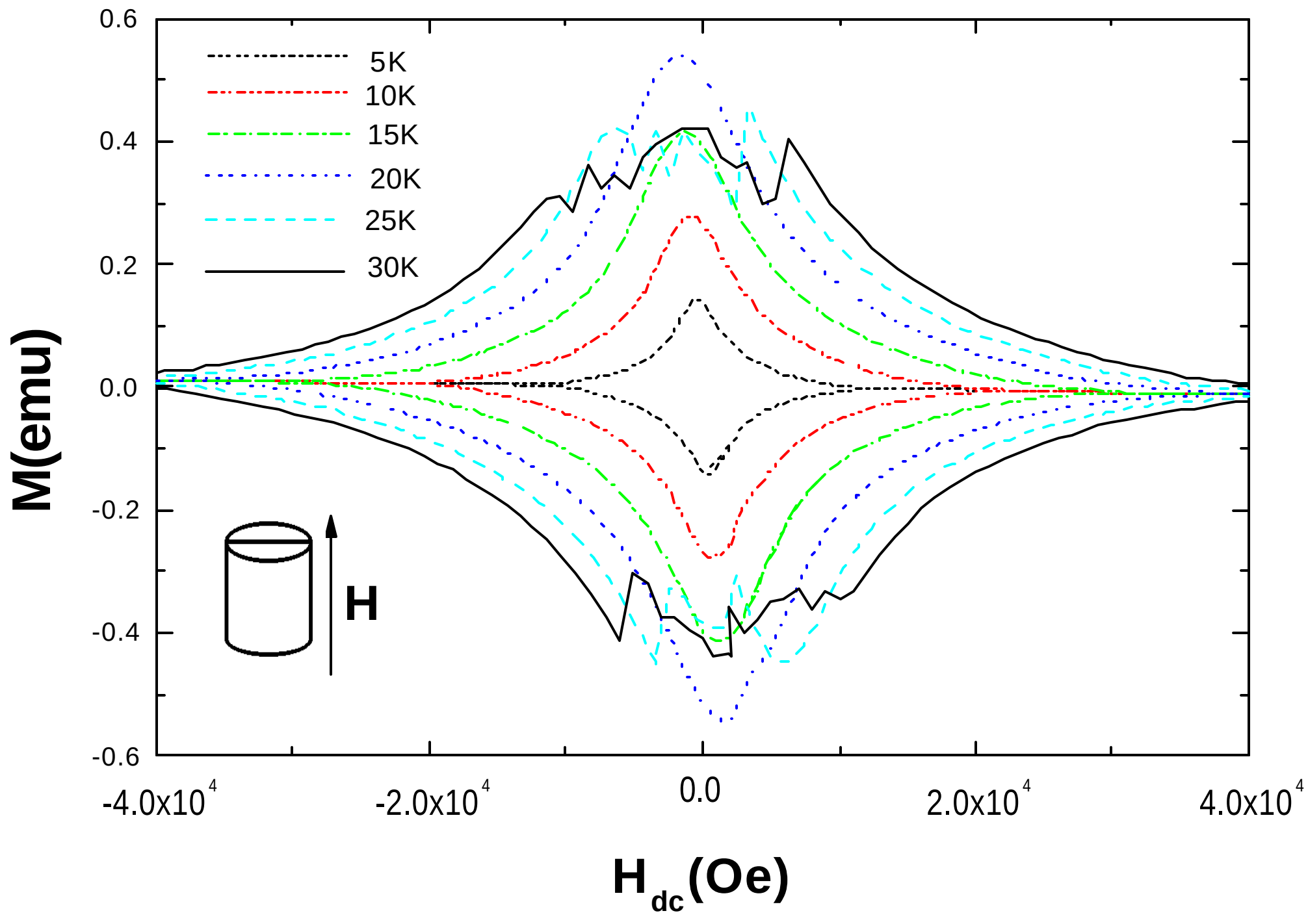

Fig. 8

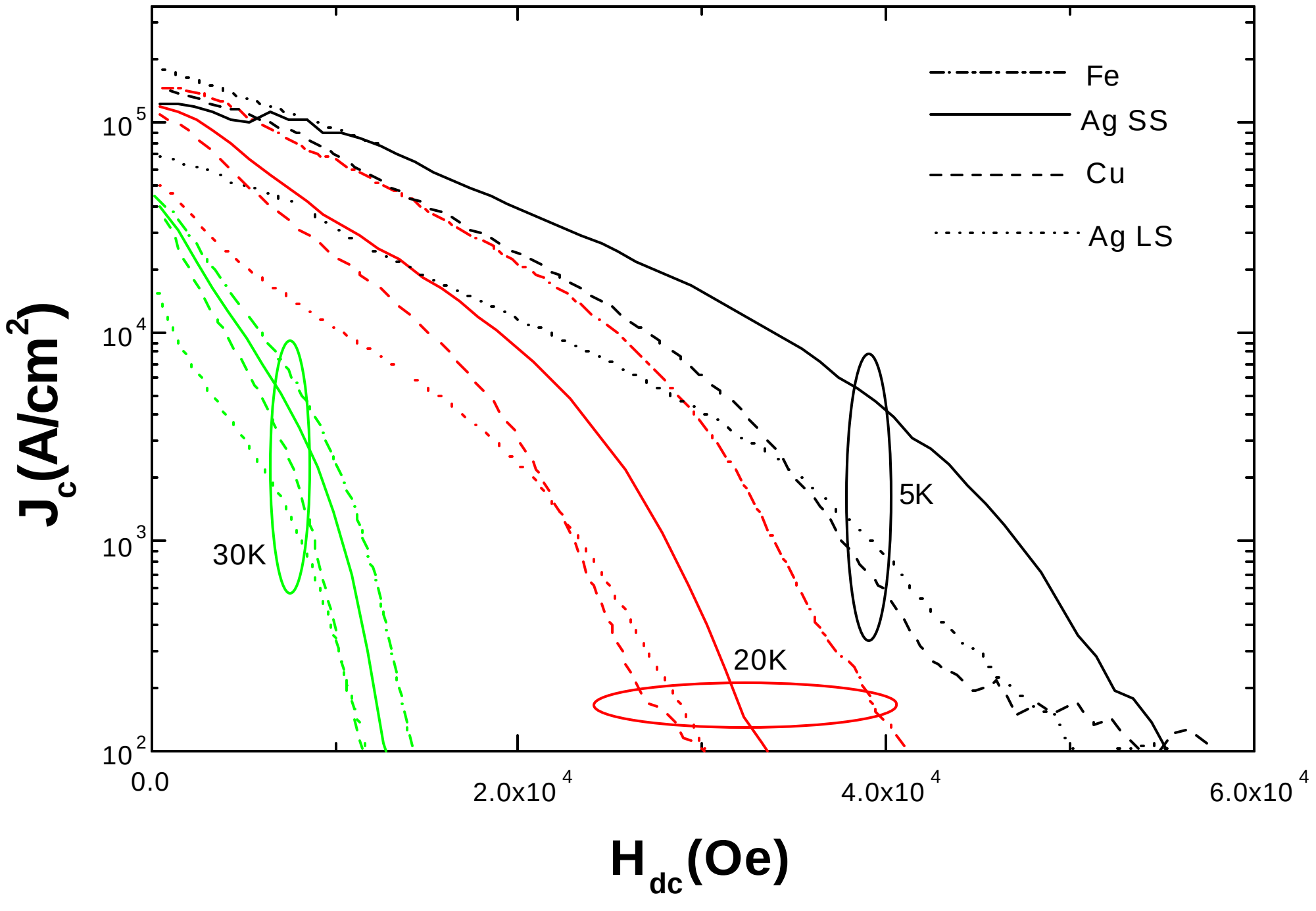

Fig. 9

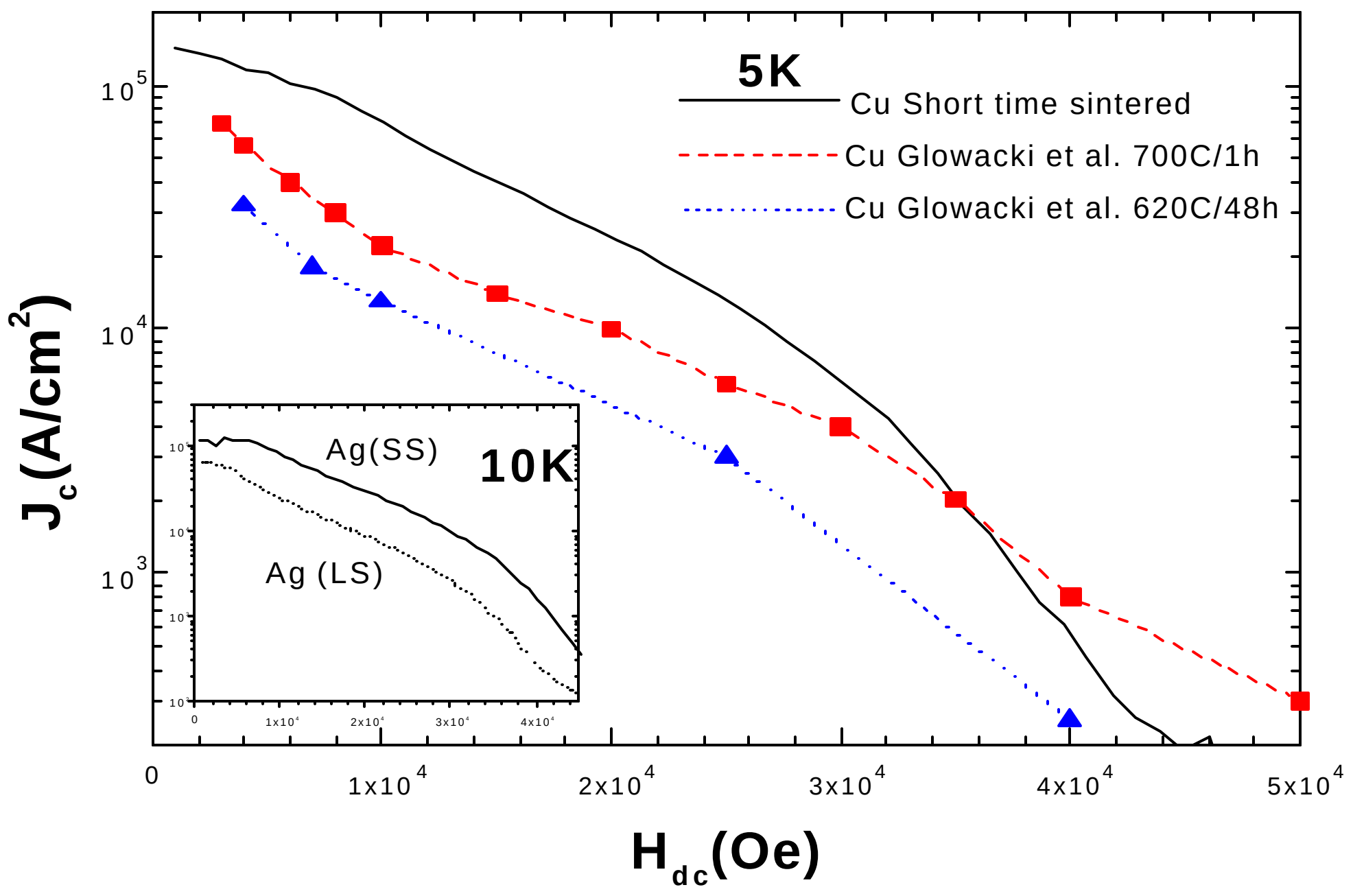
5K
Cu Short time sintered
Cu Glowacki et al. 700C/1h
Cu Glowacki et al. 620C/48h
Ag(SS)
10K
Ag (LS)
$J_c(A/cm^2)$
$H_{dc}(Oe)$

Fig. 10

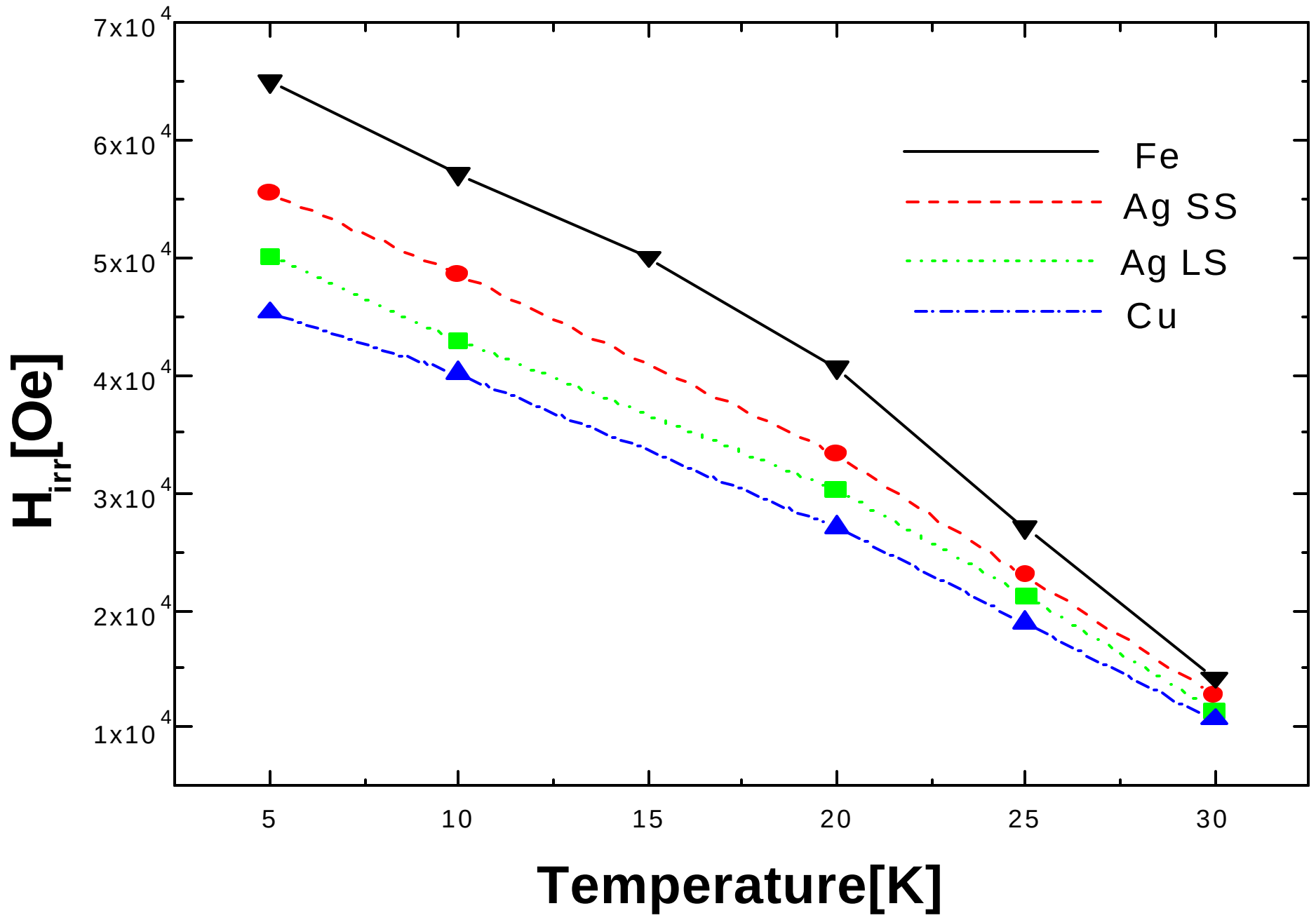

Fig. 11